\documentclass[12pt]{article}
\usepackage{amssymb}
\usepackage{amsfonts}
\usepackage{amsmath,calrsfs}
\usepackage{bbm}

\usepackage[dvips]{epsfig}
\usepackage[utf8]{inputenc}

\def\bea{\begin{eqnarray}}
\def\eea{\end{eqnarray}}
\newcommand{\bn}{\begin{eqnarray}}
\newcommand{\en}{\end{eqnarray}}
\newcommand{\nn}{\nonumber}

\def \S {\mathbb{S}}
\def \G {\mathbb{G}}
\def \i {\int{d^{3}x}}

\newcommand{\no}{\noindent}

\def\bea{\begin{eqnarray}}
\def\eea{\end{eqnarray}}

\newcommand{\be}{\begin{equation}}
\newcommand{\ee}{\end{equation}}
\def\bea{\begin{eqnarray}}
\def\eea{\end{eqnarray}}

\newcommand{\p}{\partial}

\newcommand{\s}{\,\,\,\,}

\begin{document}

\title{\textbf{Unitarity of higher derivative spin-3 models}}
\author{E. L. Mendon\c ca$^{1}$\footnote{elias.leite@unesp.br}, R. Schimidt Bittencourt $^{1}$\footnote{raphael.schimidt@unesp.br} \\
	\textit{{1- UNESP - Campus de Guaratinguet\'a - DFQ} }\\
	\textit{{CEP 12516-410, Guaratinguet\'a - SP - Brazil} }\\ }
\date{\today}
\maketitle
	
	\begin{abstract}
	The fourth order in derivatives New Massive Gravity model NMG, describes a massive spin-2 particle in $D=2+1$. At the linearized level a proof of unitarity necessarily implies that the generalization to higher dimensions includes non-unitary massless spin-2 modes. The linearized version of NMG is dual to the Fierz-Pauli model FP. Here we examine the unitarity of higher derivative spin-3 models, analogues to the NMG, dual to the Singh-Hagen model SH. We find that the same kind of restriction on the dimension of the space also happens in this case, and the models are physical only in $D=2+1$.   	
		
	\end{abstract}
	\newpage
	
	\section{Introduction}
	Three dimensional Minkowski space-times exhibit many different and special features. In such spaces, higher derivative terms can be considered in free theories without implicating in ghosts in the spectrum. The well known gauge invariant topologically massive gravity TMG \cite{djt}, with a third order topological Chern-Simons term, propagates a single state of massive spin-2 mode with positive energy. Surprisingly, in this special dimension, a parity preserving and fourth order model, which includes curvature-squared terms in the non-linear level is also free of ghosts and describes a couple of massive spin $\pm 2$ particles; this is known as the new massive gravity NMG \cite{bht}. At the linear level we have demonstrated that such theory can be understood as the soldering of two linearized TMG models with opposite helicities \cite{dm2}. Besides, the unitarity of NMG has been cared out by \cite{oda} where the authors demonstrate that the saturated propagator is free of ghost if and only if $D=2+1$.
	
	String theory is a candidate for the description of quantum gravity, its spectrum contains an infinite number of massive self-interacting higher-spin particles. It is then natural to search for higher-spin extensions of general relativity and NMG, taking advantage of the geometrical formulation provided by de Wit and Freedman \cite{dWF} and by Deser and Damour \cite{deserdam}. In this sense some progress has been reached in the spin-3 and spin-4 cases \cite{bers3,bergs4,ddelm-ms3,ddelm-nge3,ddelm-s3aux}. Specially in \cite{ddelm-ms3} we have obtained fourth and sixth order in derivatives parity preserving models describing massive spin $\pm 3$ particles, which we have argued in terms of symmetries and by counting degrees of freedom to be free of ghosts. A demonstration of unitarity for these models, as the one cared out by \cite{oda} for the case of NMG, which tell us if there is some restriction on the dimension of the space is the main subject of the present work.
	
	A consensus on what would be the  spin-3 analog of NMG has been discussed in \cite{ddelm-s3aux}, there in the sixth order in derivatives action, the auxiliary fields are not necessary, and the sixth order term is invariant under reparametrization and Weyl symmetries much like the fourth order spin-2 counterpart, named the $K$ term in \cite{bht}. However, differently of the spin-2 case, such model can not be obtained from the second order fundamental SH model as the linearized NMG can be obtained from the second order fundamental FP model, as we have demonstrated in \cite{SD_4}. On the other hand, the fourth and the sixth order models obtained in \cite{ddelm-ms3} do need the presence of auxiliary fields to remove lower spin modes, but the sixth order term by its turn is invariant under traceless reparametrizations and transverse Weyl symmetries. In fact in consequence of this difference of symmetries, we have argued that this is the reason in one model we do not need auxiliary fields while in the other we do.
	
	In order to obtain the propagator of the massive spin-3 theory we have used a spin-projection operator's basis which is quite similar to the one suggested by Barnes and Rivers to the case of rank-two fields. Once the models are gauge invariant we provide gauge-fixing terms which allow us to invert the sandwiched operator when the lagrangian is put in bilinear form. By adding source terms and consequently saturating the propagator, we have studied if the unitarity of the model can somehow survive on dimensions different then the planar world. In order to obtain the higher order lagrangians in arbitrary dimensions we have used the Noether Gauge Embedment approach NGE applied to the SH model in $D$ dimensions.

	\section{Nother gauge embedding in D dimensions}
    The SH model for massive spin-3 particles is free of ghosts in $D$ dimensions. The coefficients of the auxiliary action are dimension dependents, and can be determined even through the equations of motion, as the authors have showed in \cite{deseryang}, as also through a choice of proper spin-projection and transition operators as we have verified in \cite{elmrs}. Using scalar auxiliary fields, it is given by:  
	
	\bea
	S_{SH} =\int d^Dx \Big[\frac{1}{2} h_{\mu\nu\rho}\G^{\mu\nu\rho}(h) - \frac{m^2}{2}( h_{\mu\nu\rho} h^{\mu\nu\rho} - 3 h_{\mu} h^{\mu}) - m h_{\mu} \p^{\mu}W \Big] + S_{1} [W], \label{SH}
	\eea
	where the auxiliary action $S_{1} [W]$ is written as:
	\bea
	S_{1} [W] = \int d^Dx \Big[ \left( \frac{D}{D-2} \right)^2 m^2 W^2 -  \frac{2(D-1)}{3(D-2)} W\Box W\Big].
	\eea
	Along of this work we use the mostly plus metric $(-,+,...,+)$,  the spin-3 field is described in terms of totally symmetric rank-3 tensors
	$h_{\mu\nu\alpha}$. Besides, there are some ``geometrical'' objects named the Einstein and Schouten tensors
	which are respectively given by: \bea
	\mathbb{G}_{\mu\nu\alpha}=\mathbb{R}_{\mu\nu\alpha}-\frac{1}{2}\eta_{(\mu\nu}\mathbb{R}_{\alpha)}\quad,\quad \S^{\mu\nu\alpha}(h) \equiv \G^{\mu\nu\alpha}(h) - \frac{1}{(D+1)} \eta^{(\mu\nu}\G^{\alpha)}(h);\label{g2}\eea
	\no where we have used the spin-3 Ricci tensor and its vector contraction first introduced in
	\cite{deserdam}, namely:
	\bea \mathbb{R}_{\mu\nu\alpha}&=&\square{h}_{\mu\nu\alpha}-\partial^{\beta}\partial_{(\mu}h_{\nu\alpha)\beta}+\partial_{(\mu}\partial_{\nu}h_{\alpha)},\label{r1}\\
	\mathbb{R}_{\alpha}&=&\eta^{\mu\nu}\mathbb{R}_{\mu\nu\alpha}=2\square{h}_{\alpha}-2\partial^{\beta}\partial^{\lambda}h_{\beta\lambda\alpha}+\partial_{\alpha}\partial^{\beta}h_{\beta}. \label{r2}\eea
	
	\no 
	
	The massless spin-3 second order in derivatives term is invariant under traceless reparametrizations:
	
	\be \delta_{\tilde{\xi}}h_{\mu\nu\rho}=\p_{(\mu}\tilde{\xi}_{\nu\rho)}\label{tr}\ee
	
	\no with $\eta^{\nu\rho}\tilde{\xi}_{\nu\rho}=0$ and $\tilde{\xi}_{\nu\rho}=\tilde{\xi}_{\rho\nu}$, which is breaking by the mass term. One can impose (\ref{tr}) to the whole lagrangian using the NGE approach in order to obtain a gauge invariant fourth-order in derivatives action. In order to do that, from (\ref{SH}) we derive the Euler tensor:
	\bea
	K^{\mu\nu\rho} \equiv \frac{\delta S_{SH}}{\delta h_{\mu\nu\rho}}= \G^{\mu\nu\rho}(h) - m^2(h^{\mu\nu\rho} - \eta^{(\mu\nu}h^{\rho)}) - \frac{m}{3}\eta^{(\mu\nu}\p^{\rho)}W.
	\eea
	It is also convenient to keep in hand its trace, which is given by:
	\bea
	K^{\rho} = \G^{\rho}(h) + (D+1)m^2 h^{\rho} - \frac{(D+2)}{3}m \p^{\rho}W.
	\eea
	Introducing an extra auxiliary field $a_{\mu\nu\rho}$ with the specific gauge symmetry  $\delta_{\tilde{\xi}}a_{\mu\nu\rho}=-\delta_{\tilde{\xi}}h_{\mu\nu\rho}$,
	we implement the first iteration:
	\bea
	S_{1} = S_{SH} + \int d^Dx \;\; a_{\mu\nu\rho} K^{\mu\nu\rho}. \label{EQ1}
	\eea
	In (\ref{EQ1}) we now take the $\tilde{\xi}$-gauge variation which leave us with the following result:
	\bea
	\delta_{\tilde{\xi}} S_{1} = \int d^Dx \;\; \delta_{\tilde{\xi}} \left[  \frac{m^2}{2}(a_{\mu\nu\rho} a^{\mu\nu\rho} - 3 a_\mu a^\mu) \right], 
	\eea
	which by construction allows us to determine the second iterated action automatically $\tilde{\xi}$-gauge invariant given by:
	\bea
	S_{2} = S_{SH} + \int d^Dx \;\;  \left[ a_{\mu\nu\rho} K^{\mu\nu\rho} - \frac{m^2}{2}(a_{\mu\nu\rho} a^{\mu\nu\rho} - 3 a_\mu a^\mu) \right]. 
	\eea
	Solving the equations of motion for the auxiliary fields $a_{\mu\nu\rho}$, one can invert it in terms of the
	Euler tensors, which then give us:
	\bea
	S_{2} = S_{SH} + \frac{1}{2m^2} \int d^Dx \;\;  \left[ K_{\mu\nu\rho} K^{\mu\nu\rho} - \frac{3}{(D+1)} K_\mu K^\mu \right], \label{AX}
	\eea
	by substituting back the Euler tensor and its trace in the expression (\ref{AX}), we finally have the fourth order
	model in $D$ dimensions:
	\bea
	S^{(4)} = \int d^Dx \left[ -\frac{1}{2}h_{\mu\nu\rho}\G^{\mu\nu\rho}(h) +\frac{1}{2m^2}\G_{\mu\nu\rho}(h)\S^{\mu\nu\rho}(h) + \frac{1}{3(D+1)m}h_{\mu\nu\rho}\G^{\mu\nu\rho}(\eta\p W)\right] + S_{2}[W],\nn\\ \label{MD4}
	\eea
	where
	\bea
	S_{2}[W] = \int d^Dx \;\; \left[\left(\frac{D}{D-2}\right)^2 m^2 W^2 - \frac{D}{2(D+1)(D-2)} W\Box W\right].
	\eea
	
	One can check that our result of \cite{ddelm-ms3} can be recovered by doing $D=2+1$. The question now is if the unitarity persits in the fourth order model even in arbitrary dimensions. In order to answer this question, in the next section we carry out an analysis of the propagator of (\ref{MD4}) in $D$ dimensions.
	
	\section{Unitarity of the fourth order doublet model}
	In this section we verify the particle content of the fourth order doublet model obtained in (\ref{MD4}), after integrating over the auxiliary fields $W$ we obtain a non-local lagrangian which can be written in the bilinear form. Rewriting it in terms of the spin-projection operators given at the appendix we have:
	\bea
	\mathcal{L}^{(4)} &=& \frac{ h_{\mu\nu\alpha}}{2}\left\lbrace 
	\frac{\Box}{m^2}\left[ (\Box-m^2)P^{(3)}_{11} + Dm^2 P^{(1)}_{11} \right] \right. \nn\\
	&+& \left.
	\frac{\Box}{m^2}\left[\frac{(\Box+cm^2)}{2c}\left(3(D-1) P^{(0)}_{11} +  P^{(0)}_{22} +
	\sqrt{3(D-1)}\left( P^{(0)}_{12} + P^{(0)}_{21} \right)  \right) \right]
	\right. \nn\\
	&-& \left.
	\frac{\Box^3}{2c m^2 \omega}\left(3(D-1) P^{(0)}_{11} +  P^{(0)}_{22} +
	\sqrt{3(D-1)}\left( P^{(0)}_{12} + P^{(0)}_{21} \right)  \right) 
	\right\rbrace^{\mu\nu\alpha}_{\beta\lambda\sigma}  h^{\beta\lambda\sigma}. \label{lpro1}
	\eea
	Where we have defined $\omega=(\Box - cm^2)$ and $c=2(D+1)/(D-2)$.
	
	Once the fourth order doublet model  is invariant under the gauge transformation (\ref{tr}) we need gauge fixing terms to obtain the propagator. In order to fix the traceless reparametrizations, we suggest a de-Donder-like combination given by:
	\bea
	\mathcal{L}_{GF1} = \frac{1}{2\lambda_1} \left\lbrace  \p^\mu h_{\mu\nu\alpha} - \frac{1}{(D+2)} \left[
	\p_{( \nu} h_{\alpha )} + \eta_{\nu\alpha} (\p \cdot h) \right] \right\rbrace ^2, \eea
	
	\no where $\lambda_1$ is a gauge fixing parameter and $\p\cdot h = \p_{\mu}h^{\mu}$. This is exactly the same gauge fixing term we have used in \cite{ddelm-s3aux} in $D=2+1$. By its turn it can be rewritten in  terms of the spin-projection operators as:
	\bea
	\mathcal{L}_{GF1} &=& \frac{1}{2\lambda_1} h_{\mu\nu\alpha} \left\lbrace -\frac{\Box}{3} P_{11}^{(2)}-\frac{2}{3}\frac{(D+1)}{(D+2)^2} \Box \left[ P_{11}^{(1)} + (D+1)  P_{22}^{(1)} \right]   \right.\nn \\
	&+& \left.  \frac{2}{3}\frac{(D+1)^{3/2}}{(D+2)^2} \Box \left(  P_{12}^{(1)} + P_{21}^{(1)}  \right)   \right.\nn \\
	&+&  \left. \frac{\Box}{(D+2)^2} \left[ -3D P_{11}^{(0)} - D(D-1)
	P_{22}^{(0)} \right]
	\right.\nn \\
	&+&  \left. \frac{\Box}{(D+2)^2} D\sqrt{3(D-1)} \left(  P_{12}^{(0)} + P_{21}^{(0)}  \right)
	\right\rbrace^{\mu\nu\alpha}_{\beta\lambda\sigma}  h^{\beta\lambda\sigma}. \eea

	\no Then, one can rewrite the lagrangian (\ref{lpro1}) in a bilinear form as: 
	
	\bea	\mathcal{L}= \mathcal{L}^{(4)} + \mathcal{L}_{GF1} =h_{\mu\nu\alpha} \,\,G^{\mu\nu\alpha}_{\beta\lambda\sigma} \,\,h^{\beta\lambda\sigma}, \eea
	where the sandwiched operator $G^{\mu\nu\alpha}_{\beta\lambda\sigma}$  omitting the indices for
	sake of simplicity is given by:	
	\bea
	G &=& \frac{\Box (\Box-m^2)}{m^2} P_{11}^{(3)} - \frac{\Box}{3\lambda_1} P_{11}^{(2)} + \left[ \frac{3D(D+2)^2\lambda_1 - 2(D+1)}{3(D+2)^2\lambda_1}\right] \Box P_{11}^{(1)} \nn \\
	&-& \frac{2}{3}\left( \frac{D+1}{D+2} \right)^2 \frac{\Box}{\lambda_1} P_{22}^{(1)} + \frac{2}{3}\frac{(D+1)^{3/2}}{(D+2)^2} \frac{\Box}{\lambda_1} \left(  P_{12}^{(1)} + P_{21}^{(1)} \right)   \nn \\
	&-&  \frac{3\Box\left\lbrace D(D-2)\Box - (D+1)m^2 [2D-(D-1)(D+2)^2 \lambda_1]\right\rbrace }{(D-2)(D+2)^2(\Box - cm^2)\lambda_1} P_{11}^{(0)}  \nn \\
	&-&
	\frac{\Box\left\lbrace D(D-1)(D-2)\Box - (D+1)m^2 [2D(D-1)-(D+2)^2 \lambda_1]\right\rbrace }{(D-2)(D+2)^2(\Box - cm^2)\lambda_1} P_{22}^{(0)} \nn \\
	&+& \frac{\sqrt{3(D-1)}\Box\left\lbrace D(D-2)\Box - (D+1)m^2 [2D+(D+2)^2 \lambda_1]\right\rbrace }{(D-2)(D+2)^2(\Box - cm^2)\lambda_1}  \left( P_{12}^{(0)} + P_{21}^{(0)} \right). 
	\eea
	Once we know the identity operator (see appendix) for symmetric rank three fields we can find the inverted operator: 
	
	\bea
	G^{-1} &=& \frac{m^2}{\Box(\Box-m^2)} P_{11}^{(3)} - \frac{3\lambda_1}{\Box} P_{11}^{(2)} + \frac{1}{D\Box} P_{11}^{(1)} \nn \\
	&-& \left[ \frac{3D(D+2)^2\lambda_1 + 2(D+1)}{2D(D+1)^2 \Box}\right] P_{22}^{(1)} + \frac{\sqrt{D+1}}{D(D+1)\Box} \left(  P_{12}^{(1)} + P_{21}^{(1)} \right)  \nn \\
	&-& \frac{D(D-1)(D-2)\Box - (D+1)m^2 [2D(D-1)-(D+2)^2 \lambda_1]}{3D^3(D+1)\Box m^2} P_{11}^{(0)} \nn \\
	&-&  \frac{3\left\lbrace  D(D-2)\Box - (D+1)m^2 [2D - (D-1)(D+2)^2 \lambda_1]\right\rbrace }{3D^3(D+1)\Box m^2} P_{22}^{(0)}\nn\\
	&-& \frac{\sqrt{3(D-1)}\left\lbrace  D(D-2)\Box - (D+1)m^2 [2D + (D+2)^2 \lambda_1]\right\rbrace }{3D^3(D+1)\Box m^2} \left( P_{12}^{(0)} + P_{21}^{(0)}\right) . \nn \\
	\eea
		
	\no Now in order to analyze the physical spectrum of the model, by mean of the saturated $G^{-1}$, we consider the coupling of $h_{\mu\nu\alpha}$ to the totally symmetric source $T^{\mu\nu\alpha}$,  which give us the additional contribution: 
	\bea S &=& \int
	d^Dx \left( \mathcal{L}+ h_{\mu\nu\alpha} T^{\mu\nu\alpha} \right);\label{final} \eea 
	
	\no 
	observing that to keep the invariance under (\ref{tr})  the source must to satisfy the following restriction:
	\bea
	\delta_{\widetilde{\xi}}S=0 &\Longrightarrow& \p_\mu T^{\mu\nu\alpha} - \frac{1}{D} \eta^{\nu\alpha} \p_\mu T^\mu = 0. \label{RF11}
	\eea
	
	\no Now, we are ready to take the Fourier transform of (\ref{final}) in order to analyze the propagator in the momentum
	space saturated by totally symmetric source terms respecting the constraints (\ref{RF11}). The transition amplitude is given by: 
	
	\bea
	\mathcal{A}_2(k) &=& - \frac{i}{2} T_{\mu\nu\alpha}^{*}(k)\,\, G^{-1}(k)^{\mu\nu\alpha}_{\beta\lambda\sigma} \,\,T^{\beta\lambda\sigma}(k)\\
	&=& \frac{i}{2} \frac{1}{k^2+m^2}\left[  T_{\mu\nu\alpha}^{*} T^{\mu\nu\alpha} - \frac{3}{(D+1)} T_{\mu}^{*} T^{\mu} + \frac{(D-2)}{D^2(D+1)m^2} k^{\mu} T_{\mu}^{*} \; k_{\nu} T^{\nu} \right] \nn \\
	&-& \frac{i}{2} \frac{1}{k^2} \left[  T_{\mu\nu\alpha}^{*} T^{\mu\nu\alpha} - \frac{3}{D} T_{\mu}^{*} T^{\mu} \right] \label{24}\eea
	
	\no Following the lines of \cite{ddunitarity} we have physical particles if the imaginary part of the residue is positive i.e.: $Im\left[ Res
	(\mathcal{A}_2(k))\mid_{k^2=pole} \right] > 0$, otherwise we have a ghost in the spectrum.
	
	Let us start by the massive pole analysis, which allow us to choose the convenient rest frame where
	$k_\mu=(m,0,0,...,0)$ , which makes the equation (\ref{RF11}) becomes: \bea T^{0\nu\alpha} -
	\frac{1}{D} \eta^{\nu\alpha} T^{0} &=& 0. \eea
	In consequence we have:
	\bea
	T^{0\nu\alpha} &=& 0 \label{E1} \s\s\s\s\s\s\s\s\s \nu\neq\alpha \\
	T^{000} &=& - \frac{1}{D} T^{0} \label{E2} \\
	T^{0jj} &=&   \frac{1}{D} T^{0} \label{E3}\s\s\s\s\s j=1,2,..,(D-1). \eea
	\no Taking these information in account and noticing that $i,j,k,l,m,n=1,2,...,(D-1)$, the terms of (\ref{24}) can be rewritten as:
	\bea
	T_{\mu\nu\alpha}^{*} T^{\mu\nu\alpha} &=& T_{000}^{*} T^{000} + T_{ijk}^{*} T^{ijk} + 3T_{0ll}^{*} T^{0ll} + 3T_{00i}^{*} T^{00i}
	+ 6 T_{0mn}^{*} T^{0mn} \s\s (m\neq n) \nn \\
	&=& |T^{ijk}|^2 + \frac{(-3D+2)}{D^2}|T^0|^2 \\
	T_{\mu}^{*} T^{\mu} &=& - |T^0|^2 + |T^i|^2 \\
	k^{\mu} T_{\mu}^{*} \; k_{\nu} T^{\nu} &=& m^2 |T^0|^2
	\eea
	\no Then the imaginary part of the residue of the transition amplitude for the massive pole is given by:
	\bea
	Im\left[ Res (\mathcal{A}_2(k))\mid_{k^2=-m^2} \right] &=& \lim\limits_{k^2 \xrightarrow{} -m^2} (k^2+m^2) \mathcal{A}_2(k) \nn \\
	&=& \frac{1}{2}\left[ |T^{ijk}|^2 - \frac{3}{(D+1)} |T^{i}|^2 \right]. \eea
	Once the spatial part of the totally symmetric source $T^{ijk}$ can be decomposed into its traceless $(T^t)^{ijk}$ and trace-full $T^i$ parts as:
	\bea
	T^{ijk} = (T^t)^{ijk} + \frac{1}{(D+1)}\left(\delta^{ij}T^k + \delta^{jk}T^i + \delta^{ki}T^j \right) \s\s\s D\geq 3
	\eea
	where $\delta_{ij}(T^t)^{ijk}=0$ and $\delta_{ij}T^{ijk}=T^k$; we have:
	\bea
	|T^{ijk}|^2 = |(T^t)^{ijk}|^2 + \frac{3}{(D+1)}|T^{i}|^2 \s\s\s D\geq 3
	\eea
	Then
	\bea
	Im\left[ Res (\mathcal{A}_2(k))\mid_{k^2=-m^2} \right] = \frac{1}{2} |(T^t)^{ijk}|^2 > 0 \s\s\s \mbox{for} \s\s\s D\geq 3
	\eea
	Hence, a physical massive spin-3 particle is propagating in the spectrum for $D\geq 3$. However we still have a massless pole
	which deserve an special attention, whereas it is much more subtle. In order to study this pole we suggest a decomposition of the source in terms of polarization vectors $\epsilon^{\mu}$, $\kappa^{\mu}$ and $\tilde{\kappa}^{\mu}$ mutually orthogonals given at the appendix. The totally symmetric rank-3 source expanded in terms of the more general combination of such vectors is given by:
	\begin{eqnarray}
	T^{\mu\nu\lambda}&=& a^{ijk}\,\, \epsilon^\mu_i \epsilon^\nu_j \epsilon^\lambda_k + b^{ij}\,\, \epsilon^{(\mu}_i \epsilon^\nu_j \kappa^{\lambda)} + c^{i}\,\, \epsilon^{(\mu}_i \kappa^\nu \kappa^{\lambda)} + d\,\,\kappa^\mu \kappa^\nu \kappa^\lambda + \tilde{b}^{ij}\,\, \epsilon^{(\mu}_i \epsilon^\nu_j \tilde{\kappa}^{\lambda)} + \tilde{c}^{i}\,\, \epsilon^{(\mu}_i \tilde{\kappa}^\nu \tilde{\kappa}^{\lambda)}  \nn \\
	&+&  \tilde{d}\,\, \tilde{\kappa}^\mu \tilde{\kappa}^\nu \tilde{\kappa}^\lambda+ f^{i}\,\, \epsilon^{(\mu}_i \kappa^\nu \tilde{\kappa}^{\lambda)} + g\,\, \kappa^{(\mu} \kappa^\nu \tilde{\kappa}^{\lambda)} + h \,\,\kappa^{(\mu} \tilde{\kappa}^\nu \tilde{\kappa}^{\lambda)}.
	\end{eqnarray}
	
	\no  The arbitrary coefficients $a^{ijk}$, $b^{ij}$ and $\tilde{b}^{ij}$ are totally symmetric while the parenthesis means unnormalized symmetrization. In such base the metric tensor can be rewritten as:
	\begin{eqnarray}
	\eta^{\nu\lambda}&=& \delta^{ij} \epsilon^\nu_i \epsilon^\lambda_j - \frac{1}{2} \kappa^{(\nu} \tilde{\kappa}^{\lambda)}
	\end{eqnarray}
	while the momentum is simply $k^\mu = u\kappa^\mu$, with $u$ an arbitrary constant. Then we can check that:
	\begin{eqnarray}
	k_\mu T^{\mu\nu\lambda}&=& -2u \left[ \tilde{b}^{ij} \epsilon^{\nu}_i \epsilon^\lambda_j  + \tilde{c}^{i} \epsilon^{(\nu}_i \tilde{\kappa}^{\lambda)} + \tilde{d} \tilde{\kappa}^\nu \tilde{\kappa}^\lambda + f^{i} \epsilon^{(\nu}_i \kappa^{\lambda)} + g \kappa^\nu \kappa^{\lambda} + h \kappa^{(\nu}  \tilde{\kappa}^{\lambda)} \right] , \\
	k_\mu T^{\mu}&=& -2u \left[ \tilde{b} - 4h \right] 
	\end{eqnarray}
	But notice that in consequence of the restriction imposed by the equation (\ref{RF11}) we conclude that $\tilde{c}^{i}=f^{i}=\tilde{d}=g=0$ and besides $\tilde{b}^{ij}=-2h \delta^{ij}$, $\tilde{b} = -2h (D-2)$. With this results in hand, we have:
	\begin{eqnarray}
	T_{\mu\nu\lambda}^{*} T^{\mu\nu\lambda} &=& |a^{ijk}|^2 + 12 (h b^* + h^* b), \\
	T_{\mu}^{*} T^{\mu} &=& |a^{i}|^2 + 4D (h b^* + h^* b).	
	\end{eqnarray}
	Then the imaginary part of the residue of the transition amplitude for the massless pole can be given by:
	\bea
	Im\left[ Res (\mathcal{A}_2(k))\mid_{k^2=0} \right] &=& \lim\limits_{k^2 \xrightarrow{} 0} k^2 \mathcal{A}_2(k) \nn \\
	&=& - \frac{1}{2}\left[ |a^{ijk}|^2 - \frac{3}{D} |a^{i}|^2 \right]. \eea
	This is vanishing for $D = 3$ while it becomes negative for $D \geq 4$ as also happens with the unitarity analysis of the fourth order New Massive Gravity cared out by \cite{oda}. This is an interesting generalization of the result obtained by \cite{acioly} for the case of massive spin-2 theories.	
	 
	After all, we conclude that the fourth order doublet model (\ref{MD4})  carries a massive spin $3$ particle and is free of ghosts in its spectrum if and only if $D=2+1$ dimensions. There is no reason to attempt a new round of NGE in order to obtain a sixth order doublet model in arbitrary dimension starting from (\ref{MD4}), so in the next section we analyze the unitarity of the sixth order model we have obtained in \cite{ddelm-ms3} in $D=2+1$.
	
	\section{Unitarity of the sixth order doublet model}
	In this section we verify the particle content of the sixth order model obtained in \cite{ddelm-ms3} in three dimensions. Once the fourth order term of (\ref{MD4}) is invariant under Weyl-transverse gauge symmetry $\delta_{\psi^T}h_{\mu\nu\alpha}=\eta_{(\mu\nu}\psi_{\alpha)}^T$, one can through the NGE approach, to impose such symmetry to the whole model obtaining the sixth order model given by:
	\bea
	S^{(6)}[h]&=& \i\Big[-\frac{1}{2m^{2}}\S_{\mu\nu\alpha}(h)\G^{\mu\nu\alpha}(h)+\frac{1}{2m^{4}}\S_{\mu\nu\alpha}(h)\G^{\mu\nu\alpha}[\S(h)]+\frac{1}{12m^{3}}\S_{\mu\nu\alpha}(h)\G^{\mu\nu\alpha}(\eta \p W) \Big] \nonumber \\
     &+& \i\Big[9m^2 W^2 - \frac{9}{8} W \Box W + \frac{9}{64m^2} W \Box^2 W\Big].\label{MD6}
    \eea
    Rewriting the lagrangian in terms of spin-projection operators, after integrating over the auxiliary fields, we have:
	\bea
	\mathcal{L} &=& \frac{ h_{\mu\nu\alpha}}{2}\left\lbrace 
	\frac{\Box^2}{m^4}\left[ (\Box-m^2)P^{(3)}_{11} - \frac{(\Box+8m^2)}{128}\left( 6 P^{(0)}_{11} +  P^{(0)}_{22} +
	\sqrt{6}\left( P^{(0)}_{12} + P^{(0)}_{21} \right)  \right) \right]
	\right\rbrace^{\mu\nu\alpha}_{\beta\lambda\sigma}  h^{\beta\lambda\sigma} \nn\\
	&+& \frac{ h_{\mu\nu\alpha}}{2}\left\lbrace 
	\frac{\Box^5}{128m^4\phi}\left[ 6 P^{(0)}_{11} +  P^{(0)}_{22} +
	\sqrt{6}\left( P^{(0)}_{12} + P^{(0)}_{21} \right) \right]
	\right\rbrace^{\mu\nu\alpha}_{\beta\lambda\sigma}  h^{\beta\lambda\sigma}, \label{lpro}
	\eea
	where we have defined $\phi = (\Box - 8m^2)^2 +8\Box m^2$. Once the sixth order doublet model  is invariant under a large set of gauge symmetries i.e. the traceless reparametrizations and Weyl-transverse gauge symmetries given respectively by:
	\bea
	\delta_{\tilde{\xi}}h_{\mu\nu\alpha}&=&\p_{(\mu}\tilde{\xi}_{\nu\alpha)}, \label{gauge1} \\
	\delta_{\psi^T}h_{\mu\nu\alpha}&=& \eta_{(\mu\nu}\psi_{\alpha)}^T, \label{gauge2}
	\eea we need gauge fixing terms corresponding to the gauge parameters $\tilde{\xi}_{\nu\alpha}$ which is symmetric and traceless and $\psi_{\alpha}^T$ which is transverse, in order to obtain the propagator. To gauge fixing traceless reparametrizations, we take the same term we have used in the last section, but now with $D=3$, i.e.:
	
	\bea
	\mathcal{L}_{GF1} = \frac{1}{2\lambda_1} \left\lbrace  \p^\mu h_{\mu\nu\alpha} - \frac{1}{5} \left[
	\p_{( \nu} h_{\alpha )} + \eta_{\nu\alpha} (\p \cdot h) \right] \right\rbrace ^2, \eea
	
	\no Notice that, we have constructed this term in such a way
	that it is invariant under Weyl-trasverse transformations, and it can be rewritten in  terms of
	the spin-projection operators as:
	
	\bea
	\mathcal{L}_{GF1} &=& \frac{1}{2\lambda_1} h_{\mu\nu\alpha} \left\lbrace \Box \left[ -\frac{1}{3} P_{11}^{(2)} -\frac{8}{75} P_{11}^{(1)} -\frac{32}{75} P_{22}^{(1)} + \frac{16}{75} \left(  P_{12}^{(1)} + P_{21}^{(1)}  \right) \right]   \right\rbrace^{\mu\nu\alpha}_{\beta\lambda\sigma}  h^{\beta\lambda\sigma} \nn \\
	&+& \frac{1}{2\lambda_1} h_{\mu\nu\alpha} \left\lbrace \Box \left[ -\frac{9}{25} P_{11}^{(0)} -\frac{6}{25}
	P_{22}^{(0)} + \frac{3\sqrt{6}}{25} \left(  P_{12}^{(0)} + P_{21}^{(0)}  \right) \right]
	\right\rbrace^{\mu\nu\alpha}_{\beta\lambda\sigma}  h^{\beta\lambda\sigma}. \eea

	Once the model is still gauge invariant under Weyl-transverse transformations we add a second gauge fixing term
	given by: \bea \mathcal{L}_{GF2} = \frac{1}{2m^6\lambda_2} f_\alpha f^\alpha \label{gf2}\eea 
	
	\no with the transverse combination $f_{\alpha}$ given by: \bea f_\alpha =
	\Box\tilde{f}_\alpha - \p_\alpha(\p\cdot \tilde{f}) \quad; \quad \tilde{f}_\alpha = \p^\mu\p^\nu
	h_{\mu\nu\alpha} - \Box h_{\alpha}, \eea
	
	\no Notice that (\ref{gf2}) by its turn is invariant under traceless reparametrizations, and in terms of the
	spin-projection operators it can be written as: \bea \mathcal{L}_{GF2} &=& \frac{1}{2m^6\lambda_2}
	h_{\mu\nu\alpha} \left\lbrace \Box^4 \left[ \frac{4}{3} P_{11}^{(1)} - \frac{1}{3} P_{22}^{(1)} \right]
	\right\rbrace^{\mu\nu\alpha}_{\beta\lambda\sigma}  h^{\beta\lambda\sigma}. \eea 
	
	\no Then considering the two gauge fixing terms, one can rewrite the lagrangian (\ref{lpro}) in a bilinear form: \bea
	\mathcal{L}= \mathcal{L}^{(6)}+\mathcal{L}_{GF1}+\mathcal{L}_{GF2}= h_{\mu\nu\alpha} \,\,G^{\mu\nu\alpha}_{\beta\lambda\sigma} \,\,h^{\beta\lambda\sigma} \eea
	where the sandwiched operator $G^{\mu\nu\alpha}_{\beta\lambda\sigma}$ can be rewritten omitting the indices for
	sake of simplicity as:	
	\bea
	G &=& \frac{\Box^2(\Box-m^2)}{m^4} P_{11}^{(3)} - \frac{\Box}{3\lambda_1} P_{11}^{(2)} + \frac{4\Box}{75m^6} \left[ \frac{25\lambda_1\Box^3 - 2\lambda_2m^6}{\lambda_1\lambda_2}\right] P_{11}^{(1)} \nn \\
	&-& \frac{\Box}{75m^6} \left[ \frac{25\lambda_1\Box^3 + 32\lambda_2m^6}{\lambda_1\lambda_2}\right] P_{22}^{(1)} + \frac{16\Box}{75\lambda_1} \left[ P_{12}^{(1)} + P_{21}^{(1)} \right]  \nn \\
	&-&  \frac{3\Box(3\Box^2+200\lambda_1 \Box m^2 - 24\Box m^2 + 192 m^4)}{25\lambda_1(\Box^2 - 8 \Box m^2 + 64m^4)} P_{11}^{(0)}  \nn \\
	&-&
	\frac{2\Box(3\Box^2+50\lambda_1 \Box m^2 - 24\Box m^2 + 192 m^4)}{25\lambda_1(\Box^2 - 8 \Box m^2 + 64m^4)} P_{22}^{(0)} \nn \\
	&+& \frac{\sqrt{6} \Box(3\Box^2-100\lambda_1 \Box m^2 - 24\Box m^2 + 192 m^4)}{25\lambda_1(\Box^2 - 8 \Box m^2 + 64m^4)}  \left[ P_{12}^{(0)} + P_{21}^{(0)} \right] 
	\eea
	Once we know the identity operator for symmetric rank three fields we can find the inverted operator: \bea
	G^{-1} &=& \frac{m^4}{\Box^2(\Box-m^2)} P_{11}^{(3)} - \frac{3\lambda_1}{\Box} P_{11}^{(2)} + \frac{3m^6}{20\Box^4} \left[ \frac{(25\lambda_1\Box^3 + 32\lambda_2m^6)\lambda_2}{5\lambda_1\Box^3+6\lambda_2m^6}\right] P_{11}^{(1)} \nn \\
	&-& \frac{3m^6}{5\Box^4} \left[ \frac{(25\lambda_1\Box^3 - 2\lambda_2m^6)\lambda_2}{5\lambda_1\Box^3+6\lambda_2m^6}\right] P_{22}^{(1)} + \frac{12m^{12}}{5\Box^4}  \left[ \frac{(\lambda_2)^2}{5\lambda_1\Box^3+6\lambda_2m^6} \right]\left[  P_{12}^{(1)} + P_{21}^{(1)} \right]  \nn \\
	&-& \frac{2(3\Box^2+50\lambda_1 \Box m^2 - 24\Box m^2 + 192 m^4)}{324\Box^2 m^2} P_{11}^{(0)} \nn \\
	&-& \frac{3(3\Box^2+200\lambda_1 \Box m^2 - 24\Box m^2 + 192 m^4)}{324\Box^2 m^2} P_{22}^{(0)}\nn\\
	&-& \frac{\sqrt{6}(3\Box^2-100\lambda_1 \Box m^2 - 24\Box m^2 + 192 m^4)}{324\Box^2 m^2}  \left[ P_{12}^{(0)} + P_{21}^{(0)} \right]. \nn \\
	\eea

	Now in order to analyze the physical spectrum of the model we consider again the coupling of $h_{\mu\nu\alpha}$ to the
	totally symmetric source term $T^{\mu\nu\alpha}$,  which give us the additional contribution: \bea S &=& \int
	d^3x \left( \mathcal{L} + h_{\mu\nu\alpha} T^{\mu\nu\alpha} \right); \eea \no observing that to keep the
	invariance under (\ref{gauge1}) and (\ref{gauge2}) the source term must to satisfy the following restrictions:
	\bea
	\delta_{\widetilde{\xi}}S=0 &\Longrightarrow& \p_\mu T^{\mu\nu\alpha} - \frac{1}{3} \eta^{\nu\alpha} \p_\mu T^\mu = 0, \label{RF1}\\
	\delta_{\psi^T}S=0 &\Longrightarrow& T^\mu = \p^\mu \varOmega. \label{RF2} \eea
	
	\no Where $\Omega$ is an arbitrary scalar function. Now, we are ready to take the Fourier transform of the previous result in order to analyze the propagator in the momentum
	space saturated by totally symmetric source terms respecting the constraints (\ref{RF1}) and (\ref{RF2}). Then we look to the
	imaginary part of the residue of the transition amplitude given by: \bea
	\mathcal{A}_2(k) &=& - \frac{i}{2} T_{\mu\nu\alpha}^{*}(k)\,\, G^{-1}(k)^{\mu\nu\alpha}_{\beta\lambda\sigma} \,\,T^{\beta\lambda\sigma}(k)\\
	&=& \frac{i}{2} \frac{1}{k^2+m^2}\left[  T_{\mu\nu\alpha}^{*} T^{\mu\nu\alpha} - \frac{7}{9} k^2 \varOmega^2 \right] - \frac{i}{2}\frac{1}{k^2}\left[  T_{\mu\nu\alpha}^{*} T^{\mu\nu\alpha} \right] \nn \\
	&+& \frac{i}{2} \frac{m^2}{k^4} \left[  T_{\mu\nu\alpha}^{*} T^{\mu\nu\alpha} + k^2 \varOmega^2 \right] +
	\frac{7i}{18}\varOmega^2 + \frac{i}{2} \frac{(k^2 + 8m^2)}{36m^2}\varOmega^2 \label{at6}\eea
	
	Such result is quite similar to the one we have found in \cite{ddelm-s3aux} for a sixth order doublet model which do not depend on auxiliary fields. In fact, we can observe that, the effect of the auxiliary fields on the present model is codified in the last term of (\ref{at6}), which do not affect the particle content of the model. Then we can state that the transition amplitude (\ref{at6}) is also free of ghosts propagating a unique massive spin-3 particle. In addition notice that in \cite {ddelm-s3aux} the set of symmetries which keeps the sixth order term invariant are larger than here, once in the present case the gauge parameters have restrictions such as the tracelessness and the traversity, while there they are full. In a certain way, we see that the role of the auxiliary fields has been played by the increase of symmetries in the sixth order term \cite{ddelm-s3aux}.
	
	\section{Conclusion}
	Starting with the second order Singh-Hagen model for massive spin-3 particles in $D$ dimensions (\ref{SH}), we have imposed through the NGE approach invariance under traceless reparametrizations, obtaining then a gauge invariant fourth order model (\ref{MD4}). Integrating over the auxiliary fields we have put the lagrangian (\ref{MD4}) in bilinear form and rewriting it in terms of spin projection operators. Once it is gauge invariant under traceless reparametrizations, a de Donder like gauge-fixing term is considered in order to obtain the propagator. By adding a source term we analyze the saturated transition amplitude, concluding that the massive spin-3 mode is propagating only if $D\geq 3$. On the other hand for the massless pole, the imaginary part of the transition amplitude is vanishing if $D=3$ and is negative if $D>3$, which implies in a ghost for such dimensions. The result is completely equivalent with the one obtained by \cite{oda} when studying the unitarity of NMG for the spin-2 case, and also is generalization for rank-3 fields of the work cared out by \cite{acioly}.
	
	The sixth order model (\ref{MD6}) is invariant under traceless reparametrizations $\delta_{\tilde{\xi}}h_{\mu\nu\lambda}=\p_{(\mu}\tilde{\xi}_{\nu\lambda)}$ and transverse Weyl transformations $\delta_{\psi^T}h_{\mu\nu\lambda}=\eta_{(\mu\nu}\psi^T_{\lambda)}$. It can be obtained from the fourth order model as we have demonstrated at \cite{ddelm-ms3,ddelm-nge3}, which implies in a duality relation between the SH theory and the fourth and sixth order models. Due to the invariance under transverse Weyl transformations we have considered another gauge fixing term in order to invert the sandwiched operator. The addition of the extra gauge-fixing term is done in such a way that it preserves invariance under traceless reparametrizations, as well as the first gauge-fixing is invariant under transverse Weyl transformations. Proceeding with the analysis of the saturated amplitude one can easy see that, except by the last term in (\ref{at6}), which do not interfere in the particle content, it is in fact the same we have obtained in \cite{ddelm-s3aux}. Then one can state that the model is free of ghosts propagating only a massive spin-3 particle.
	
	Analogies with the NMG are also commented, we argue that although the recent sixth order model we have obtained in \cite{ddelm-s3aux} is free of ghosts, with no need of auxiliary fields and have the highest order term invariant under full reparametrizations and Weyl transformations as it is the case of the K-term of \cite{bht}, it can not be obtained from the SH theory as the NMG can be obtained from the Fierz-Pauli theory. On the other hand, the doublet models we have proved unitarity here comes from the SH model even through the master action technique as from the NGE approach, we say that they are dual to the SH theory. In fact the master action approach can demonstrates such duality only if one assume that the massless second order term is free of particle content, what happens only in $D=2+1$ dimensions, in total agreement with what we have verified here.

	\section{Appendix}
	
	Taking the spin-1 and spin-0 projection operators $\theta_{\mu\nu}=\eta_{\mu\nu}-\omega_{\mu\nu}$ and
	$\omega_{\mu\nu}={\p_{\mu}\p_{\nu}}/{\Box}$, one can construct in $D$ dimensions, the spin-3 projection
	operators as follows:

	\bea
	(P^{(3)}_{11})^{\mu\nu\rho}_{\alpha\beta\gamma} &=&  \theta^{(\mu}_{(\alpha} \theta^\nu_\beta \theta^{\rho)}_{\gamma)} - (P^{(1)}_{11})^{\mu\nu\rho}_{\alpha\beta\gamma},\label{first} \\
	(P^{(2)}_{11})^{\mu\nu\rho}_{\alpha\beta\gamma} &=&  3 \theta^{(\mu}_{(\alpha} \theta^\nu_\beta \omega^{\rho)}_{\gamma)} - (P^{(0)}_{11})^{\mu\nu\rho}_{\alpha\beta\gamma},  \\
	(P^{(1)}_{11})^{\mu\nu\rho}_{\alpha\beta\gamma} &=&  \frac{3}{(D+1)}  \theta^{(\mu\nu} \theta_{(\alpha\beta}  \theta^{\rho)}_{\gamma)},\\
	(P^{(1)}_{22})^{\mu\nu\rho}_{\alpha\beta\gamma} &=&  3 \theta^{(\mu}_{(\alpha} \omega^\nu_\beta \omega^{\rho)}_{\gamma)},  \\
	(P^{(0)}_{11})^{\mu\nu\rho}_{\alpha\beta\gamma} &=&  \frac{3}{(D-1)}  \theta^{(\mu\nu} \theta_{(\alpha\beta}  \omega^{\rho)}_{\gamma)}, \label{p110} \\
	(P^{(0)}_{22})^{\mu\nu\rho}_{\alpha\beta\gamma} &=&   \omega_{\alpha\beta}
	\omega^{\mu\nu} \omega^{\rho}_{\gamma} \label{p220} \eea

	\no Notice that, here the parenthesis means normalized symmetrization, taking for example the first term in (\ref{first}) we have: \be \theta^{(\mu}_{(\alpha} \theta^\nu_\beta \theta^{\rho)}_{\gamma)}=\frac{1}{6}(
	\theta^{\mu}_{\alpha} \theta^\nu_\beta \theta^{\rho}_{\gamma}+
	\theta^{\rho}_{\alpha} \theta^\nu_\beta \theta^{\mu}_{\gamma}+
	\theta^{\nu}_{\alpha} \theta^\mu_\beta \theta^{\rho}_{\gamma}+
	\theta^{\rho}_{\alpha} \theta^\mu_\beta \theta^{\nu}_{\gamma}+
	\theta^{\nu}_{\alpha} \theta^\rho_\beta \theta^{\mu}_{\gamma}+
	\theta^{\mu}_{\alpha} \theta^\rho_\beta \theta^{\nu}_{\gamma}).\ee   
	The totally symmetric identity operator is represented by $\mathbbm{1}$
	and is given by: \bea \mathbbm{1}^{\mu\nu\rho}_{\alpha\beta\gamma} = 
	\delta^{(\mu}_{(\alpha} \delta^\nu_\beta \delta^{\rho)}_{\gamma)}. \label{id3} \eea Finally, the transition operators $P^{(s)}_{{ij}}$
	are given by: \bea
	(P^{(1)}_{{12}})^{\mu\nu\rho}_{\alpha\beta\gamma} &=&  \frac{3}{\sqrt{(D+1)}}   \theta_{(\alpha\beta} \theta^{(\rho}_{\gamma)} \omega^{\mu\nu)},  \\
	(P^{(1)}_{{21}})^{\mu\nu\rho}_{\alpha\beta\gamma} &=&  \frac{3}{\sqrt{(D+1)}}   \theta^{(\mu\nu} \theta^{\rho)}_{(\gamma} \omega_{\alpha\beta)},  \\
	(P^{(0)}_{{12}})^{\mu\nu\rho}_{\alpha\beta\gamma} &=&  \frac{3}{\sqrt{3(D-1)}}   \theta_{(\alpha\beta} \omega^{(\mu\nu} \omega^{\rho)}_{\gamma)},   \\
	(P^{(0)}_{{21}})^{\mu\nu\rho}_{\alpha\beta\gamma} &=& 
	\frac{3}{\sqrt{3(D-1)}}   \theta^{(\mu\nu} \omega_{(\alpha\beta} \omega^{\rho)}_{\gamma)}. \eea \no

	In terms of the basis $(\epsilon,\kappa,\tilde{\kappa})$ a vector field could be decomposed as follows:
	\begin{eqnarray}
	A^\mu = a \kappa^\mu + b \tilde{\kappa}^\mu + t^i \epsilon^\mu_i \s\s\s\s i=1,2,...,(D-2)
	\end{eqnarray}
	where $a$ , $b$ and $t^i$ are expansion coefficients, while:
	\begin{eqnarray}
	\kappa^\mu&=&(1,0,0,....,0,1),  \\
	\tilde{\kappa}^\mu&=&(1,0,0,....,0,-1),  \\
	\epsilon^0&=&(0,0,0,....,0,0),  \\
	\epsilon^1&=&(0,1,0,....,0,0),  \\
	\epsilon^2&=&(0,0,1,....,0,0), \s\s\s\s \mbox{until ...}  \\
	\epsilon^{(D-2)}&=&(0,0,0,....,1,0),  \\
	\epsilon^{(D-1)}&=&(0,0,0,....,0,0). 
	\end{eqnarray} 
	Notice that the basis is linearly independent, with inter products given by:
	\begin{eqnarray}
	\kappa_\mu \epsilon^\mu &=& 0, \nn \\
	\tilde{\kappa}_\mu \epsilon^\mu &=& 0, \nn \\
	\kappa_\mu \kappa^\mu &=& 0, \nn \\
	\tilde{\kappa}_\mu \tilde{\kappa}^\mu &=& 0, \nn \\
	\kappa_\mu \tilde{\kappa}^\mu &=& -2. \nn
	\end{eqnarray}

	\section{Acknowledgements}
	
	This study was financed in part by the Coordenação de Aperfeiçoamento de Pessoal de Nivel Superior - Brasil (CAPES) - Finance Code 001. 
	We acknowledge helpful discussions with Prof. Denis Dalmazi.

\end{document}